\documentclass[aps,english,showpacs,twocolumn,prl]{revtex4-1}
\usepackage{amsfonts}
\usepackage{amssymb}
\usepackage{amsmath}
\usepackage{graphicx}
\usepackage{epsfig}
\usepackage{color,xcolor}
\usepackage{hyperref}

\hypersetup{hypertex=true,colorlinks=true,linkcolor=blue,anchorcolor=blue,citecolor=blue}

\begin{document}

\title{Robust Incoherent Perfect Absorption}
\author{H. S. Xu}
\author{L. Jin}
\email{jinliang@nankai.edu.cn}
\affiliation{School of Physics, Nankai University, Tianjin 300071, China}

\begin{abstract}
Coherent perfect absorber is capable of completely absorbing input waves.
However, the coherent perfect absorption severely depends on the
superposition of the input waves, and the perfect absorption is sensitive to
the disorder of the absorber. Thus, a robust incoherent perfect absorption,
being insensitive to the superposition of input waves and the system
disorder, is desirable for practical applications. Here, we demonstrate that
the linearly independent destructive interference at the port connections
removes the constraint on the coherent input. We propose a novel approach
using the interplay between the loss and localization to form the incoherent
perfect absorption. The resonant incidence from either port is completely
absorbed. Furthermore, we utilize the lattice configuration that supporting
the flat band to demonstrate the disorder-immune incoherent perfect
absorption. Our findings provide insight into the fundamentals and
applications for the perfect absorption of light, microwave, sound,
mechanical wave and beyond.
\end{abstract}

\maketitle

\textit{Introduction}.---Dissipation ubiquitously exists in physical
systems. Unlike the conventional wisdom to suppress dissipation,
non-Hermitian optics actively engineers the dissipation and has stimulated
many intriguing phenomena that never been found in Hermitian systems.
Especially, the interplay of interference and dissipation enables the
complete absorption of light with the proper coherent inputs \cite%
{ADStonePRL10,ADStoneSci11,LonghiPRA12,HChenPRL14,YDChongNRM17,JQYouNC17,VVKSA18,JeffersPRl19,KottosNC20,KottosPRR20,JQYouPRA20,YLaiLight21,WJWanPhotoniX22,CWQiuPRB23,BinkowskiPRB24}%
. This phenomenon, known as coherent perfect absorption, serves as the
time-reversed counterpart of lasing \cite%
{MostafazadehPRL09,LonghiPRA10,StonePRL12,LFengNP16,LFengPRA17,DAZPRA19,VVKPRA19,LonghiOL19,VVKOL20,XHSPRA21,XHSPRA23}%
. The coherent perfect absorption has immediate applications in
optical switch, modulator, and imaging for the coherent control of waves
\cite{ZheludevAPL14,XFangLSA15,LiewAP16,AluPRX16}. However, the proper
superposition of input waves is a prerequisite to achieve the complete
absorption. This rigorous constraint on the coherent input limits the
absorption efficiency and hinders the practical application of perfect
absorber \cite{FaccioNC15,FaccioPRL16}. To overcome the restriction,
unidirectional perfect absorber is designed and the input from one side is
completely absorbed \cite%
{RamezaniPRL14,LonghiOL15,JLPRL18,JLSR16,RamezaniIJTQE16,SHFanNP17}.
Furthermore, bidirectional perfect absorber is capable of completely
absorbing resonant input from either side \cite%
{SHFanAPL14,YLaiEPL16,SQChenAOM17,KatzScience22}. The perfect absorption of
incoherent input regardless of the superposition of incident waves is
desirable. However, its design is a touch task. In addition, the perfect
absorption is also sensitive to the disorder in the scattering center, any
fabrication imperfection may cause the deviation from the perfect absorption.

In this Letter, we find that the linearly independent destructive
interference of the degenerate eigenstates of the effective scattering
center at the port connections plays an important role to overcome the
constraint of perfect absorption on the coherent input, and we propose a
novel approach to realize incoherent perfect absorption. In addition, we
demonstrate the disorder-immune incoherent perfect absorption created from
the compact localization of the flat band in a stub ribbon of coupled
resonators. The incident waves from both ports at any superposition are
completely absorbed in the presence of random coupling disorder. The
proposed models can also be implemented in many experimental platforms. The
formalism is applicable for many other flat band lattices. Our findings open
up novel avenue in the design of highly efficient absorbers for practical
applications.

\begin{figure*}[tb]
\includegraphics[bb=0 0 296 40, width=17.8 cm, clip]{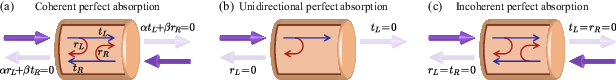}
\caption{Schematic of three types of perfect absorptions. (a) Coherent
perfect absorption: the reflection and transmission vanish for coherent
input waves from the opposite directions. The perfect absorption highly
depends on the relative amplitudes and phases of input waves. (b)
Unidirectional perfect absorption: the reflection and transmission vanish
for input wave from one direction. (c) Incoherent perfect absorption: the
output vanish for any superposition of input waves from the opposite
directions. $r_L$, $t_L$ ($r_R$, $t_R$) are the reflection and
transmission for the left (right) input.}
\label{fig1}
\end{figure*}

\textit{Perfect absorption.---}In a two-port scattering system, each
incident wave is split into a reflected wave and a transmitted wave after
interference in the scattering center. Interestingly, the losses embedded in
a non-Hermitian scattering center may completely absorb the input waves. In
Fig.~\ref{fig1}, we illustrate three types of perfect absorptions. The
coherent perfect absorber realizes complete absorption of resonant input
waves from both ports at a proper superposition of the wave amplitudes and
phases [Fig.~\ref{fig1}(a)]. The coherent perfect absorber is highly
sensitive to the variation of input waves. The unidirectional perfect
absorber realizes complete absorption of input wave from one port [Fig.~\ref%
{fig1}(b)]. The unidirectional perfect absorber does not require the proper
superposition of input waves, but the transmission and/or reflection for the
wave injected from another port is nonzero \cite%
{LonghiOL15,JLPRL18,JLSR16,RamezaniIJTQE16}. Interestingly, the incoherent
perfect absorber realizes complete absorption of input wave from either port
[Fig.~\ref{fig1}(c)]. The proper match of input wave amplitudes and phases
is unnecessary, and any superposition of the resonant inputs from both ports
can be completely absorbed. The degenerate resonances enable
incoherent perfect absorber~\cite{SHFanAPL14}.

The complete absorption of input indicates that the steady-state solution of
the perfect absorber possesses the purely incoming waves at the real
frequency. Notably, both the coherent perfect absorber, which requires a
coherent input, and the unidirectional perfect absorber, which requires a
specific input direction, have the non-degenerate purely incoming wave
solution. By contrast, as a result of the complete absorption of incoming
waves from both the left and right ports, the incoherent perfect absorber
has two degenerate purely incoming wave solutions at a real frequency. This
differs from the two purely incoming wave solutions coalesced to a single
state \cite%
{StonePRL19,LYangScience21,StonePRA22,Ozdemir22,HougnePRL22,TKottosCP22,HougneNC22,SakoticLPR23}%
, where the absorption exhibits quartic line shape. The former one is at the
diabolic point of the non-Hermitian system and the later one is at the
exceptional point of the non-Hermitian system. For the incoherent perfect
absorption, the incoming wave solutions from each port are linearly
independent. This overcomes the constraint on the relative amplitude and
phase of the input waves at the corresponding frequency.

The physical isolation between the two ports to overcome the
constraint on the coherent input is not under our
consideration, i.e., two decoupled single-port perfect absorbers formally
combined as a two-port system apparently realizes an incoherent perfect
absorption. Here we highlight the interference effect in the scattering
center rather than the physical isolation. Two necessary conditions are
required to form incoherent perfect absorption. (i) The input frequency is
the degenerate energy level of the effective scattering center, which is the
scattering center with additional on-site terms reduced from the ports at
the incoming wave solution. (ii) The degenerate eigenstates are able to form
destructive interference at the connections, where the scattering center
connected with the ports. Then, the incoming wave is confined within each
port and its extension feature is destroyed.

In our formalism, the linearly independent destructive interference at the
port connections is important for the incoherent perfect absorption, i.e., two on-resonance degenerate eigenstates of the effective
scattering center respectively have vanishing amplitudes at the two port
connections. The microscopic origin of the degenerate resonant states differs in
distinct physical systems \cite{SHFanAPL14}.
Notably, the flat band is fully constituted by degenerate states. The flat
band eigenstates are compact localized within one or several unit cells of
the lattice. The wave input on-resonance with the flat band energy cannot
propagate through the lattice. Thus, the unit cells of flat band lattice as
the scattering center are excellent candidates to generate incoherent
perfect absorption if the reflection can be perfectly absorbed via
engineering the losses.

\begin{figure}[tb]
\includegraphics[bb=0 0 380 385, width=8.8 cm, clip]{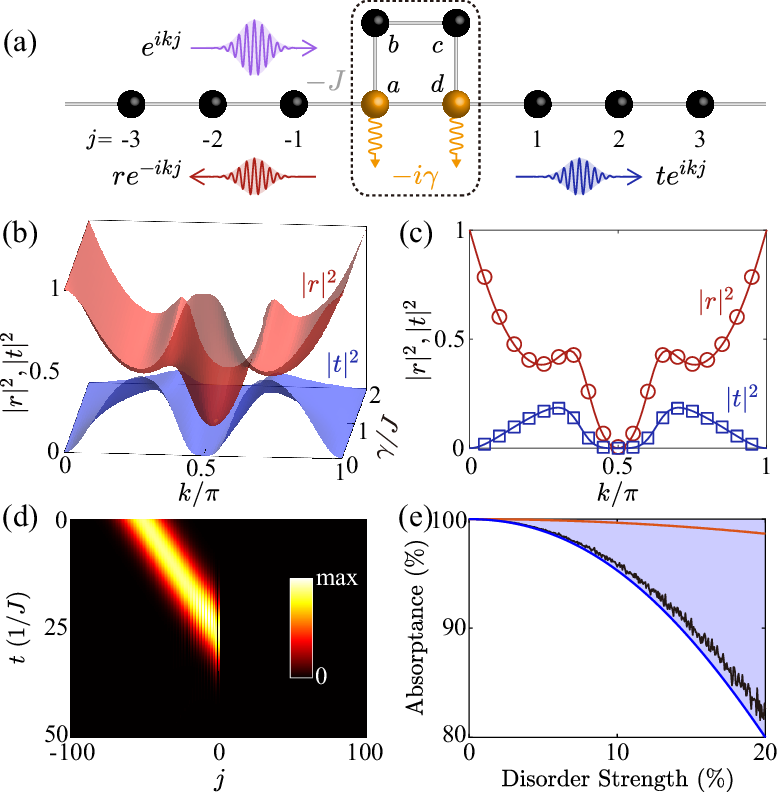}
\caption{(a)
Schematic of a non-Hermitian square plaquette (dashed black box) embedded in the resonator
array. (b) The reflection (in red) and transmission (in blue) as a function of
the input wave vector and loss. (c) The reflection and transmission at $\protect\gamma=J$. (d) The simulations of perfect absorption for the initial Gaussian wave packet centered at $N_{\mathrm{c}}=-50$ with $\protect\sigma=10$ and $k_{\mathrm{c}}=\protect\pi/2$. (e) The absorptance as a function of disorder strength. The blue curve is the analytic result of the minimum absorptance, and the black (red) curve is the numerical result of the minimum (average) absorptance of $10^5$ samples. The disorder is randomly chosen with a uniform distribution in the interval ($-1,1$) of disorder strength. The shaded area is the disordered absorptance range.}
\label{fig2}
\end{figure}

\textit{Non-Hermitian square plaquette.---}We consider a uniform chain of
evanescently coupled resonators [Fig.~\ref{fig2}(a)] to demonstrate the
formation of incoherent perfect absorption. A non-Hermitian square plaquette
is embedded in the middle of the resonator chain at two neighbor dissipative
resonators, the resonators have the loss $-i\gamma $. The couplings are $-J$%
. The resonators have the resonant frequency $\omega _{\mathrm{c}}$ and
support the counterclockwise and clockwise modes, which are the
time-reversal counterparts \cite{LeykamPRB19}. The scattering properties for
the counterclockwise and clockwise mode are the same. In the coupled mode
theory \cite{JLPRL18,StonePRA20,XHSPRR22}, the equation of motion for the
resonator mode of the ports is
\begin{equation}
i\frac{\mathrm{d}f_{j}}{\mathrm{d}t}=-Jf_{j-1}-Jf_{j+1},(\left\vert
j\right\vert >1),  \label{EOM0a}
\end{equation}%
and
\begin{equation}
i\frac{\mathrm{d}f_{\mp 1}}{\mathrm{d}t}=-Jf_{\mp 2}-Jf_{a(d)},
\label{EOM0b}
\end{equation}%
where $f_{j}$ ($\left\vert j\right\vert \geq 1$) is the mode amplitude for
the port resonator $j$ and $f_{s}$ ($s=a,b,c,d$) is the mode amplitude for
the scattering center resonator $s$. The resonant frequency term $\omega _{%
\mathrm{c}}$ is removed from the equations of motion for simplicity. The
system is equivalently transformed into a rotating frame with the frequency $%
\omega _{\mathrm{c}}$. The equations of motion for the resonator modes of
the scattering center are%
\begin{eqnarray}
i\frac{\mathrm{d}f_{a}}{\mathrm{d}t} &=&-Jf_{d}-Jf_{b}-Jf_{-1}-i\gamma f_{a},
\label{EOM1} \\
i\frac{\mathrm{d}f_{b}}{\mathrm{d}t} &=&-Jf_{a}-Jf_{c},  \label{EOM2} \\
i\frac{\mathrm{d}f_{c}}{\mathrm{d}t} &=&-Jf_{b}-Jf_{d},  \label{EOM3} \\
i\frac{\mathrm{d}f_{d}}{\mathrm{d}t} &=&-Jf_{c}-Jf_{a}-Jf_{1}-i\gamma f_{d}.
\label{EOM4}
\end{eqnarray}

In the elastic scattering process, we have $f_{j(s)}=\psi _{j(s)}e^{-i\omega
\left( k\right) t}$, where $\psi _{j(s)}$ is the steady-state wave function
of the resonator. The incoming wave is reflected and transmitted by the
scattering center. The steady-state solution is a superposition of two
counter-propagating plane waves $e^{\pm ikj}$, where the wave momentum $k$
is a real number. For the wave injected from the left side of the resonator
array, we denote the steady-state wave function in the left port as $\psi
_{j}=e^{ikj}+re^{-ikj}$ ($j<0$) and the steady-state wave function in the
right port as $\psi _{j}=te^{ikj}$ ($j>0$), where $r$ and $t$ are the
reflection and transmission coefficients, respectively. The scattering
center has the inversion symmetry. Thus, the reflection and transmission are
symmetric \cite{LJinCPL21,SHFanPRL22}, being independent of the direction of
input wave.

Substituting the wave functions into Eqs.~(\ref{EOM0a}) and (\ref%
{EOM0b}), we obtain the dispersion $\omega (k)=-2J\cos k$ of the ports.
Substituting the wave functions into Eqs.~(\ref{EOM1})-(\ref{EOM4}), we
obtain the scattering coefficients (see Supplemental Material A \cite{SI}).
The reflection and transmission as the functions of the input wave momentum $%
k$ are depicted in Figs.~\ref{fig2}(b) and \ref{fig2}(c). The total
intensity of the scattered light is less than unity as a consequence of the
losses \cite{XHSPRR23}. At the resonant input $k=\pi /2$ for $\gamma =J$,
the perfect absorption occurs with vanishing reflection and transmission $%
r=t=0$. In this case, the scattering matrix is a $2\times 2$ null matrix.
Thus, any superposition of resonant input wave is perfectly absorbed at the
non-Hermitian square plaquette, forming a incoherent perfect absorption. In
Fig.~\ref{fig2}(d), we perform the dynamics of the incoherent perfect
absorption for the resonant wave injected in the left port \cite%
{GaussianWavePacket}. The numerical simulations well agree with the
predictions.

Incoherent perfect absorption is independent of the relative phases and
amplitudes of the incoming waves from both ports and overcomes the
limitation of perfect absorption being confined to coherent or
unidirectional input. Nevertheless, we emphasize that the incoherent perfect
absorption is still sensitive to the system parameters and the disorder in
the scattering center affects the absorption efficiency. Then, the incident
waves cannot be completely absorbed. Figure \ref{fig2}(e) shows the
decreased absorptance as a function of disorder strength in the couplings of
non-Hermitian square plaquette. The minimum absorptance is $80\%$ at a
coupling disorder strength of $20\%$ (see Supplemental Material B \cite{SI}%
). The random disorder causes fluctuation in the observed minimum
absorptance, which reduces as the number of sample increases.

\textit{Effective scattering center.---}We consider the purely incoming wave
solution of the two-port non-Hermitian scattering system shown in Fig. \ref%
{fig2}(a)
\begin{equation}
\Psi _{j}=\left\{
\begin{array}{c}
\alpha e^{ikj},(j<0) \\
\beta e^{-ikj},(j>0)%
\end{array}%
\right. ,  \label{PIS}
\end{equation}%
where $\alpha :\beta $ indicates the relative amplitude and phase of the
input waves in the left and right ports. It is important to note that any
superposition of input waves from the left and right ports can be completely
absorbed, making any form of $\alpha :\beta $ a perfect absorption solution
for the incoherent perfect absorber. In the following analysis, the momentum
$k$ does not necessarily have to be a real number in $\Psi _{j}$.

The projection theory provides an effective Hamiltonian that equivalently
describes the original system in a reduced dimension \cite{JLPRA10,JLPRA11},
which is the effective scattering center. From Eq. (\ref{PIS}), the wave
function continuity yields the wave functions of the connection resonator $a$
and $d$ are $\Psi _{a}=\alpha e^{ik\cdot 0}$ and $\Psi _{d}=\beta
e^{-ik\cdot 0}$. From Eqs.~(\ref{EOM1}) and (\ref{EOM4}), the contribution
of the wave function at the port to the scattering center is $-J\alpha
e^{ik\cdot (-1)}$ and $-J\beta e^{-ik\cdot 1}$. From the equivalence $%
-J\alpha e^{ik\cdot (-1)}\equiv -Je^{-ik}\alpha e^{ik\cdot 0}=-Je^{-ik}\Psi
_{a}$ and $-J\beta e^{-ik\cdot 1}\equiv -Je^{-ik}\beta e^{-ik\cdot
0}=-Je^{-ik}\Psi _{d}$, the coupling to the ports effectively reduces into
the on-site term $-Je^{-ik}$ at the resonators $a$\ and $d$. In this
scenario, the effective scattering center takes the form of
\begin{equation}
H_{\mathrm{c}}^{\mathrm{eff}}(k)=-\left(
\begin{array}{cccc}
i\gamma +Je^{-ik} & J & 0 & J \\
J & 0 & J & 0 \\
0 & J & 0 & J \\
J & 0 & J & i\gamma +Je^{-ik}%
\end{array}%
\right) .
\end{equation}%
The purely incoming wave solution usually corresponds to a complex
frequency, which is not a steady-state solution as this leads to the input
waves growing or decaying exponentially with time. However, the purely
incoming wave solution can be turned to a real frequency by the loss,
forming the steady-state of perfect absorption.

\begin{figure}[tb]
\includegraphics[bb=0 0 375 185, width=8.8 cm, clip]{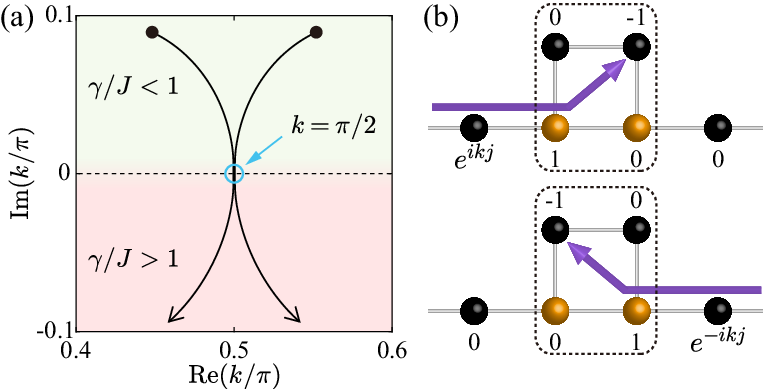}
\caption{The
purely incoming solution of the non-Hermitian square plaquette of
Fig.~\ref{fig2}(a). (a) Trajectories of two $k$ for the purely incoming
solution as $\gamma$ varies from $\gamma/J=0$ (black solid circles) to
$\gamma/J=2$ (black arrows), and they meet at $\gamma/J=1$ (cyan hollow
circle). (b) Two degenerate eigenstates of the non-Hermitian square
plaquette at $\protect\gamma/J=1$.} \label{fig3}
\end{figure}

The effective scattering center includes the energy of the input wave, i.e.,
\begin{equation}
\det |H_{\mathrm{c}}^{\mathrm{eff}}(k)-\omega (k)|=0,  \label{Jost solution}
\end{equation}%
where we obtain the momentum $k$ that satisfies the purely incoming
solution. We plot the solution for different loss rate $\gamma $ in Fig.~\ref%
{fig3}(a). When $\gamma <J$ ($\gamma >J$), there are two complex $k$ located
in the upper (lower) half-plane, i.e., $\mathrm{Im}(k)>0$ ($\mathrm{Im}(k)<0$%
). For $\gamma =J$, the two complex $k$ meet on the real axis at $k=\pi /2$.
The two perfect absorption solutions are degenerate and linearly independent
as verified from the effective scattering center $H_{\mathrm{c}}^{\mathrm{eff%
}}(\pi /2)$, which has two degenerate energy levels at $\omega (\pi /2)=0$.
The corresponding degenerate eigenstates are $(1,0,-1,0)^{T}$ and $%
(0,-1,0,1)^{T}$, as illustrated in Fig.~\ref{fig3}(b). Two degenerate
eigenstates constitute a pair of inversion symmetric counterparts as a
consequence that the scattering center is inversion symmetric. 
In this case, the inversion symmetry plays an important role in the
formation of incoherent perfect absorption~\cite{SHFanAPL14}, which is not
necessary in a more general framework. This point is demonstrated as follows.

The destructive interference of on-resonance degenerate eigenstates from the
effective scattering center at single connection resonator provides a
strategy to destroy the extension feature of the perfectly absorbing state
and create the incoherent perfect absorption. The localization on different
ports caused by the destructive interference ensures that the incoming waves
from each port are completely absorbed (see Supplemental Material C \cite{SI}%
). Specifically, the non-zero wave function in the resonator $a$ ($d$)
represents the incoming wave from the left (right) port, while the zero wave
function in the resonator $d$ ($a$) causes the vanishing of incoming wave
from the right (left) port. Therefore, the scattering center independently
absorbs the incoming waves from the left and right ports.

\begin{figure}[tb]
\includegraphics[bb=0 0 387 234, width=8.8 cm, clip]{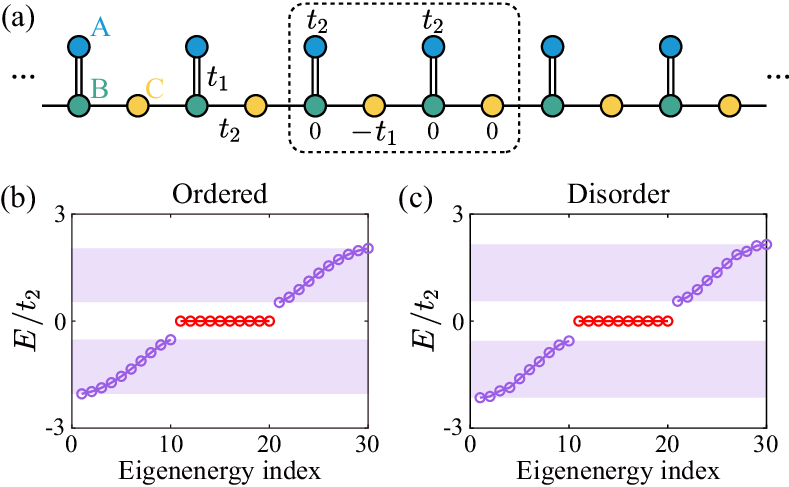}
\caption{(a) Schematic of the stub ribbon. The sublattices $A$, $B$, $C$ are denoted by
the blue, green, and yellow solid circles, respectively. The dashed box depicts one of the
compact localized states associated with the zero-energy flat band. (b) and
(c) are the eigenenergies of a $30$-site stub ribbon without disorder and
with disorder, respectively. The parameter is chosen $t_1=t_2/2$.
In (c), all the couplings deviate from the set parameters within the range
of [$-20\%$, $20\%$].} \label{fig4}
\end{figure}

\textit{Stub ribbon}.---The incoherent perfect absorber is not unique. The
identification of effective scattering center holding degenerate eigenstates
becomes a critical point to obtain an incoherent perfect absorber. The
flat-band lattice meets the prerequisite for the incoherent perfect
absorption. Notably, the dispersionless flat band holds a huge number of
degenerate energy levels \cite%
{LeykamAPX18,LeykamAPL18,ZGChenNanophotonics20,ZGChenPRL20,ZGChenAPL21,PobleteAPX21}%
. These eigenstates compactly localize within several neighbor unit cells
across the lattice \cite%
{FlachEL14,GneitingPRB18,ZSMPRB20,ZSMPRA23}. These compact
localize states are linear independent. Thus, the flat-band lattice is a
promising candidate for an effective scattering center. The incoherent
perfect absorption is possible through engineering the losses.

\begin{figure*}[tb]
\includegraphics[bb=0 0 283 52, width=17.8 cm, clip]{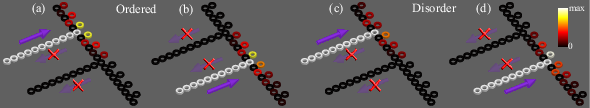}
\caption{The steady-state solutions for different incident directions [(a),
(b)] without disorder and [(c), (d)] with disorder. The color bar indicates
the intensity of wave function. The non-Hermitian stub ribbon is
the stub ribbon in Fig.~\protect\ref{fig4} with additional losses at the two
connection sites. The port coupling is chosen $J=t_2$.}
\label{fig5}
\end{figure*}

We consider a stub ribbon in Fig.~\ref{fig4}(a) \cite{LeykamPRB17}. The stub
ribbon is a prototypical quasi-one-dimensional flat-band lattice composed of
three sublattices ($A$, $B$, $C$) \cite%
{LiGePRA15,BlochPRL16,LiGePR18,LQYuanAP22}. The sublattices have the
resonant frequency $\omega _{\mathrm{c}}$. The lattice is formed by the
coupling $t_{1}$ between the sublattices $A$ and $B$ and the coupling $t_{2}$
between the sublattices $B$ and $C$. The Hamiltonian of the stub ribbon is
\begin{equation}
H_{\mathrm{stub}}=\sum_{n}[t_{1}A_{n}^{\dag }B_{n}+t_{2}(B_{n}^{\dag
}C_{n}+C_{n}^{\dag }B_{n+1})]+\mathrm{H.c.},
\end{equation}%
where $A_{n}^{\dag }$, $B_{n}^{\dag }$, and $C_{n}^{\dag }$ ($A_{n}$, $B_{n}$%
, and $C_{n}$) denote the creation (annihilation) operators for the three
sublattices in the $n$-th unit cell. We plot the eigenenergies of a $30$%
-site stub ribbon in Fig.~\ref{fig4}(b). The stub ribbon has two dispersive
bands and one zero-energy flat band in the middle. The flat band exhibits a
tenfold degeneracy, which increases with the expansion of the lattice size.
The wave function amplitudes for each site of the unnormalized eigenstate of
the flat band are depicted in Fig.~\ref{fig4}(a), which are compactly
localized within two neighbor unit cells and have nonvanishing amplitudes
only at the sublattices $A$ and $C$. Notably, the flat band in the stub
ribbon is protected by the chiral symmetry (see Supplemental Material D \cite%
{SI}), rendering it independent of the coupling strengths. The chiral
symmetry and flat band persist even in the presence of coupling disorder
\cite{JoglekarPRA11,DavyNature22}. The eigenenergies in the presence of
coupling disorder is shown in Fig.~\ref{fig4}(c). The coupling disorder
affects two dispersive bands, leading to the perturbations in their
eigenenergies \cite{LeykamPRB13}. While the zero-energy flat band is immune
to the coupling disorder similar as topological edge state \cite{HCWuPRL24}.
The robustness is always desirable for practical applications \cite%
{JoglekarPRA10,JoglekarPRAR11,FleuryPRL19,KottosPRApp20,ClerkPRL23}.

\textit{Robust incoherent perfect absorber.---}To design a robust incoherent
perfect absorber, the stub ribbon as the scattering center is connected to
two semi-infinite coupled resonator chains (Fig.~\ref{fig5}). The two
resonator chains serve as the input and output ports, which are coupled to
the sublattice $C$ sites from \textit{any} two different unit cells of the
stub ribbon. The port coupling is $-J$ and the resonant frequency is $\omega
_{\mathrm{c}}$.

We focus on the on-resonant incident wave with the momentum $k=\pi /2$. The
steady-state solutions formed by the incident wave injected from the upper
and lower ports into the stub ribbon are shown in Figs.~\ref{fig5}(a) and~%
\ref{fig5}(b), respectively. The losses at the connection resonators are $%
\gamma =J$. The propagation direction of the plane wave is indicated by the
purple arrow. The uniform intensity observed at the input port reveals the
presence of only incident waves, indicating the vanishing reflection.
Moreover, the zero intensity at the output port indicates the vanishing
transmission. The incident wave injected from the input port propagates
without reflection to the output port until being absorbed at the connection
resonators. The incident waves injected from different ports are both
trapped within the stub ribbon and completely absorbed without any
reflection or transmission. These demonstrate the incoherent perfect
absorption.

The proposed incoherent perfect absorber, without the requirement of
coherent input, exhibits robust high absorption efficiency. We emphasize
that the zero-energy flat band is a key factor for the formation of
incoherent perfect absorption in the stub ribbon. The superposition of
compact localized states of the flat band as the incoming wave solution of
the scattering system creates the destructive interference at the two
connection resonators, respectively. This ensures the complete absorption of
incoming wave injected from both directions. Furthermore, the flat band
maintains in the presence of random coupling disorder. Consequently, the
incoherent perfect absorption is robust against the random coupling disorder
although the distribution of the steady-state solutions within the stub
ribbon is affected. The steady-state solutions in the presence of random
coupling disorder are depicted in Figs.~\ref{fig5}(c) and~\ref{fig5}(d). In
addition, the incoherent perfect absorption induced by the stub ribbon is
also robust against the random detuning in the sublattice $B$ (see
Supplemental Material E \cite{SI}).

The proposed approach for the construction of robust incoherent perfect
absorber can be applied to other quasi-one-dimensional systems such as the
rhombic lattice \cite{SzameitNC20,AhufingerLight20,VicencioPRL22} and the
two-dimensional systems such as the Lieb lattice \cite%
{LeykamPRA12,ThomsonPRL15,VicencioPRL15}, including a various of
non-Hermitian flat-band lattices \cite%
{MolinaPRA15,RamezaniPRA17,LGePRL18,ASzameitPRL19,LJinPRA19,SMZhangPRA19,SMZhangPRR20,AAndreanovPRB21}%
. The incoherent perfect absorption is induced by the interplay between the
engineered loss and flat band localization. In addition, the incoherent
perfect absorption protected by the specific symmetries demonstrates
remarkable robustness to the disorder. The robust incoherent perfect
absorption can be implemented in the coupled waveguides and other platforms
\cite{LChenAP23}.

\textit{Conclusion.---}We have found that the linearly independent
destructive interference of the on-resonance degenerate eigenstates from the
effective scattering center at the port connections create the incoherent
perfect absorption. The resonant inputs are completely absorbed without the
requirement of a proper coherent input, which is beneficial for the future
applications. The degenerate eigenstates of the effective
scattering center are required, but the inversion symmetry is not necessary~{\color{red}\cite{SHFanAPL14}.} Furthermore, we have proposed a robust incoherent
perfect absorption using the flat band localization. The proposed incoherent
perfect absorption protected by the chiral symmetry is immune to the
disorder, leading to the high absorption efficiency. Our findings are
insightful for the perfect absorption of light, microwave, sound, mechanical
wave and beyond.

This work was supported by National Natural Science Foundation of China
(Grant No.~12222504).


\begin{thebibliography}{999}
\bibitem{ADStonePRL10} Y. D. Chong, L. Ge, H. Cao, and A. D. Stone, Coherent
Perfect Absorbers: Time-Reversed Lasers, Phys. Rev. Lett. \textbf{105},
053901 (2010).

\bibitem{ADStoneSci11} W. Wan, Y. Chong, L. Ge, H. Noh, A. D. Stone, and H.
Cao, Time-reversed lasing and interferometric control of absorption, Science
\textbf{331}, 889 (2011).

\bibitem{LonghiPRA12} S. Longhi and G. D. Valle, Coherent perfect absorbers
for transient, periodic, or chaotic optical fields: Time-reversed lasers
beyond threshold, Phys. Rev. A \textbf{85}, 053838 (2012).

\bibitem{HChenPRL14} Y. Sun, W. Tan, H.-Q. Li, J. Li, and H. Chen,
Experimental Demonstration of a Coherent Perfect Absorber with PT Phase
Transition, Phys. Rev. Lett. \textbf{112}, 143903 (2014).

\bibitem{YDChongNRM17} D. G. Baranov, A. Krasnok, T. Shegai, A. Al\`{u}, and
Y. Chong, Coherent perfect absorbers: Linear control of light with light,
Nat. Rev. Mater. \textbf{2}, 17064 (2017).

\bibitem{JQYouNC17} D. Zhang, X.-Q. Luo, Y.-P. Wang, T.-F. Li, and J. Q.
You, Observation of the exceptional point in cavity magnon-polaritons, Nat.
Commun. \textbf{8}, 1368 (2017).

\bibitem{VVKSA18} A. M\"{u}llers, B. Santra, C. Baals, J. Jiang, J. Benary,
R. Labouvie, D. A. Zezyulin, V. V. Konotop, and H. Ott, Coherent perfect
absorption of nonlinear matter waves, Sci. Adv. \textbf{4}, eaat6539 (2018).

\bibitem{JeffersPRl19} J. Jeffers, Nonlocal Coherent Perfect Absorption,
Phys. Rev. Lett. \textbf{123}, 143602 (2019).

\bibitem{KottosNC20} L. Chen, T. Kottos, and S. M. Anlage, Perfect
absorption in complex scattering systems with or without hidden symmetries,
Nat. Commun. \textbf{11}, 5826 (2020).

\bibitem{KottosPRR20} Q. Zhong, L. Simonson, T. Kottos, and R. El-Ganainy,
Coherent virtual absorption of light in microring resonators, Phys. Rev.
Res. \textbf{2}, 013362 (2020).

\bibitem{JQYouPRA20} W. Xiong, J. Chen, B. Fang, C. Lam, and J. Q. You,
Coherent perfect absorption in a weakly coupled atom-cavity system, Phys.
Rev. A \textbf{101}, 063822 (2020).

\bibitem{YLaiLight21} J. Luo, H. Chu, R. Peng, M. Wang, J. Li, and Y. Lai,
Ultra-broadband reflectionless Brewster absorber protected by reciprocity,
Light Sci. Appl. \textbf{10}, 89 (2021).

\bibitem{WJWanPhotoniX22} J. Hou, J. Lin, J. Zhu, G. Zhao, Y. Chen, F.
Zhang, Y. Zheng, X. Chen, Y. Cheng, L. Ge, and W. Wan, Self-induced
transparency in a perfectly absorbing chiral second-harmonic generator,
PhotoniX \textbf{3}, 22 (2022).

\bibitem{CWQiuPRB23} M. Liu, W. Chen, G. Hu, S. Fan, D. N. Christodoulides,
C. Zhao, and C.-W. Qiu, Spectral phase singularity and topological behavior
in perfect absorption, Phys. Rev. B \textbf{107}, L241403 (2023).

\bibitem{BinkowskiPRB24} F. Binkowski, F. Betz, R. Colom, P. Genevet, and S.
Burger, Poles and zeros in non-Hermitian systems: Application to photonics,
Phys. Rev. B \textbf{109}, 045414 (2024).

\bibitem{MostafazadehPRL09} A. Mostafazadeh, Spectral Singularities of
Complex Scattering Potentials and Infinite Reflection and Transmission
Coefficients at Real Energies, Phys. Rev. Lett. \textbf{102}, 220402 (2009).

\bibitem{LonghiPRA10} S. Longhi, $\mathcal{PT}$-symmetric laser absorber,
Phys. Rev. A \textbf{82}, 031801(R) (2010).

\bibitem{StonePRL12} Y. D. Chong, L. Ge, and A. D. Stone, $\mathcal{PT}$%
-Symmetry Breaking and Laser-Absorber Modes in Optical Scattering Systems,
Phys. Rev. Lett. \textbf{106}, 093902 (2012).

\bibitem{LFengNP16} Z. J. Wong, Y.-L. Xu, J. Kim, K. O'Brien, Y. Wang, L.
Feng, and X. Zhang, Lasing and anti-lasing in a single cavity, Nat. Photon.
\textbf{10}, 796 (2016).

\bibitem{LFengPRA17} L. Ge and L. Feng, Contrasting eigenvalue and
singular-value spectra for lasing and antilasing in a $\mathcal{PT}$%
-symmetric periodic structure, Phys. Rev. A \textbf{95}, 013813 (2017).

\bibitem{DAZPRA19} V. V. Konotop and D. A. Zezyulin, Spectral singularities
of odd-$\mathcal{PT}$-symmetric potentials, Phys. Rev. A \textbf{99}, 013823
(2019).

\bibitem{VVKPRA19} V. V. Konotop, E. Lakshtanov, and B. Vainberg, Designing
lasing and perfectly absorbing potentials, Phys. Rev. A \textbf{99}, 043838
(2019).

\bibitem{LonghiOL19} S. Longhi, Photonic flat-band laser, Opt. Lett. \textbf{%
44}, 287 (2019).

\bibitem{VVKOL20} D. A. Zezyulin and V. V. Konotop, Universal form of arrays
with spectral singularities, Opt. Lett. \textbf{45}, 3447 (2020).

\bibitem{XHSPRA21} H. S. Xu and L. Jin, Coupling-induced nonunitary and
unitary scattering in anti-$\mathcal{PT}$-symmetric non-Hermitian systems,
Phys. Rev. A \textbf{104}, 012218 (2021).

\bibitem{XHSPRA23} H. S. Xu, L. C. Xie, and L. Jin, High-order spectral
singularity, Phys. Rev. A \textbf{107}, 062209 (2023).

\bibitem{ZheludevAPL14} X. Fang, M. L. Tseng, J. Ou, K. F. MacDonald, D. P.
Tsai, N. I. Zheludev, Ultrafast all-optical switching via coherent
modulation of metamaterial absorption, Appl. Phys. Lett. \textbf{104},
141102 (2014).

\bibitem{XFangLSA15} X. Fang, K. F. MacDonald, and N. I. Zheludev,
Controlling light with light using coherent metadevices: all-optical
transistor, summator and invertor, Light Sci. Appl. \textbf{4}, e292 (2015).

\bibitem{LiewAP16} S. F. Liew, S. M. Popoff, S. W. Sheehan, A. Goetschy, C.
A. Schmuttenmaer, A. D. Stone, and H. Cao, Coherent control of photocurrent
in a strongly scattering photoelectrochemical system, ACS Photon. \textbf{3}%
, 449 (2016).

\bibitem{AluPRX16} F. Monticone, C. A. Valagiannopoulos, and A. Al\`{u},
Parity-Time Symmetric Nonlocal Metasurfaces: All-Angle Negative Refraction
and Volumetric Imaging, Phys. Rev. X \textbf{6}, 041018 (2016).

\bibitem{FaccioNC15} T. Roger, S. Vezzoli, E. Bolduc, J. Valente, J. J. F.
Heitz, J. Jeffers, C. Soci, J. Leach, C. Couteau, N. I. Zheludev, and D.
Faccio, Coherent perfect absorption in deeply subwavelength films in the
single-photon regime, Nat. Commun \textbf{6}, 7031 (2015).

\bibitem{FaccioPRL16} T. Roger, S. Restuccia, A. Lyons, D. Giovannini, J.
Romero, J. Jeffers, M. Padgett, and D. Faccio, Coherent Absorption of N00N
States, Phys. Rev. Lett. \textbf{117}, 023601 (2016).

\bibitem{RamezaniPRL14} H. Ramezani, H.-K. Li, Y. Wang, and X. Zhang,
Unidirectional Spectral Singularities, Phys. Rev. Lett. \textbf{113}, 263905
(2014).

\bibitem{LonghiOL15} S. Longhi, Non-reciprocal transmission in photonic
lattices based on unidirectional coherent perfect absorption, Opt. Lett.
\textbf{40}, 1278 (2015).

\bibitem{JLPRL18} L. Jin and Z. Song, Incident Direction Independent Wave
Propagation and Unidirectional Lasing, Phys. Rev. Lett. \textbf{121}, 073901
(2018).

\bibitem{JLSR16} L. Jin, P. Wang, and Z. Song, Unidirectional perfect
absorber, Sci. Rep. \textbf{6}, 32919 (2016).

\bibitem{RamezaniIJTQE16} H. Ramezani, Y. Wang, E. Yablonovitch, and X.
Zhang, Unidirectional Perfect Absorber, IEEE J. Sel. Top. Quantum Electron.
\textbf{22}, 115 (2016).

\bibitem{SHFanNP17} Y. Huang, Y. Shen, C. Min, S. Fan, and G. Veronis,
Unidirectional reflectionless light propagation at exceptional points,
Nanophotonics \textbf{6}, 977 (2017).

\bibitem{SHFanAPL14} J. R. Piper, V. Liu, and S. Fan, Total absorption by
degenerate critical coupling, Appl. Phys. Lett. \textbf{104}, 251110 (2014).

\bibitem{YLaiEPL16} P. Bai, Y. Wu, and Y. Lai, Multi-channel coherent
perfect absorbers, Europhys. Lett. \textbf{114} 28003 (2016).

\bibitem{SQChenAOM17} J. Li, P. Yu, C. Tang, H. Cheng, J. Li, S. Chen, and
J. Tian, Bidirectional perfect absorber using free substrate plasmonic
metasurfaces, Adv. Opt. Mater. \textbf{5}, 1700152 (2017).

\bibitem{KatzScience22} Y. Slobodkin, G. Weinberg, H. H\"{o}rner, K.
Pichler, S. Rotter, and O. Katz, Massively degenerate coherent perfect
absorber for arbitrary wavefronts, Science \textbf{377}, 995 (2022).

\bibitem{StonePRL19} W. R. Sweeney, C. W. Hsu, S. Rotter, and A. D. Stone,
Perfectly Absorbing Exceptional Points and Chiral Absorbers, Phys. Rev.
Lett. \textbf{122}, 093901 (2019).

\bibitem{LYangScience21} C. Wang, W. R. Sweeney, A. D. Stone, and L. Yang,
Coherent perfect absorption at an exceptional point, Science \textbf{373},
1261 (2021).

\bibitem{StonePRA22} A. Farhi, A. Mekawy, A. Al\`{u}, and D. Stone,
Excitation of absorbing exceptional points in the time domain, Phys. Rev. A
\textbf{106}, L031503 (2022).

\bibitem{Ozdemir22} S. Soleymani, Q. Zhong, M. Mokim, S. Rotter, R.
El-Ganainy, and S. K. \"{O}zdemir, Chiral and degenerate perfect absorption
on exceptional surfaces, Nat. Commun. \textbf{13}, 599 (2022).

\bibitem{HougnePRL22} C. Ferise, P. d. Hougne, S. F\'{e}lix, V. Pagneux, and
M. Davy, Exceptional Points of $\mathcal{PT}$-Symmetric Reflectionless
States in Complex Scattering Systems, Phys. Rev. Lett. \textbf{128}, 203904
(2022).

\bibitem{TKottosCP22} S. Suwunnarat, Y. Tang, M. Reisner, F. Mortessagne, U.
Kuhl, and T. Kottos, Non-linear coherent perfect absorption in the proximity
of exceptional points, Commun. Phys. \textbf{5}, 5 (2022).

\bibitem{HougneNC22} J. Sol, D. R. Smith, and P. del Hougne,
Meta-programmable analog differentiator, Nat. Commun. \textbf{13}, 1713
(2022).

\bibitem{SakoticLPR23} Z. Sakotic, P. Stankovic, V. Bengin, A. Krasnok, A. Al%
\`{u}, and N. Jankovic, Non-Hermitian Control of Topological Scattering
Singularities Emerging from Bound States in the Continuum, Laser Photon.
Rev. \textbf{10}, 2200308 (2023).

\bibitem{LeykamPRB19} J. Han, C. Gneiting, and D. Leykam, Helical transport
in coupled resonator waveguides, Phys. Rev. B \textbf{99}, 224201 (2019).

\bibitem{StonePRA20} W. R. Sweeney, C. W. Hsu, and A. D. Stone, Theory of
reflectionless scattering modes, Phys. Rev. A \textbf{102}, 063511 (2020).

\bibitem{XHSPRR22} H. S. Xu and L. Jin, Coherent resonant transmission,
Phys. Rev. Res. \textbf{4}, L032015 (2022).

\bibitem{LJinCPL21} L. Jin and Z. Song, Symmetry-Protected Scattering in
Non-Hermitian Linear Systems, Chin. Phys. Lett. \textbf{38}, 024202 (2021).

\bibitem{SHFanPRL22} C. Guo and S. Fan, Reciprocity constraints on
reflection, Phys. Rev. Lett. \textbf{128}, 256101 (2022).

\bibitem{SI} The Supplementary Material provide the detailed calculations of
scattering coefficients, analytical expression of the minimum absorptance,
an example of coherent perfect absorber, flat band and CLSs of stub ribbon,
and steady-state solutions in the presence of detuning.

\bibitem{XHSPRR23} H. S. Xu and L. Jin, Pseudo-Hermiticity protects the
energy-difference conservation in the scattering, Phys. Rev. Res. \textbf{5}%
, L042005 (2023).

\bibitem{GaussianWavePacket} The plane wave injection has a Gaussian profile
$|\phi \left( 0\right) \rangle =\Omega ^{-1/2}\sum_{j}e^{-(j-N_{\mathrm{c}%
})^{2}/(2\sigma ^{2})}e^{ik_{\mathrm{c}}j}\left\vert j\right\rangle $, where
$\Omega $ is the normalization factor, $\sigma $ controls the width, $N_{%
\mathrm{c}}$ is the spatial center, and $k_{\mathrm{c}}$ is the central
momentum. The Gaussian wave packet input from either side towards the
scattering center is completely absorbed without any reflection and
transmission.

\bibitem{JLPRA10} L. Jin and Z. Song, Physics counterpart of the $\mathcal{PT%
}$ non-Hermitian tight-binding chain, Phys. Rev. A \textbf{81}, 032109
(2010).

\bibitem{JLPRA11} L. Jin and Z. Song, Partitioning technique for discrete
quantum systems, Phys. Rev. A \textbf{83}, 062118 (2011).

\bibitem{LeykamAPX18} D. Leykam, A. Andreanov, and S. Flach, Artificial flat
band systems: from lattice models to experiments, Adv. Phys. X \textbf{3},
1473052 (2018).

\bibitem{LeykamAPL18} D. Leykam and S. Flach, Perspective: Photonic
flatbands, APL Photon. \textbf{3}, 070901 (2018).

\bibitem{ZGChenNanophotonics20} L. Tang, D. Song, S. Xia, S. Xia, J. Ma, W.
Yan, Y. Hu, J. Xu, D. Leykam, and Z. Chen, Photonic flat-band lattices and
unconventional light localization, Nanophotonics \textbf{9}, 1161 (2020).

\bibitem{ZGChenPRL20} J. Ma, J. Rhim, L. Tang, S. Xia, H. Wang, X. Zheng, S.
Xia, D. Song, Y. Hu, Y. Li, B. Yang, D. Leykam, and Z. Chen, Direct
Observation of Flatband Loop States Arising from Nontrivial Real-Space
Topology, Phys. Rev. Lett. \textbf{124}, 183901 (2020).

\bibitem{ZGChenAPL21} Y. Xie, L. Song, W. Yan, S. Xia, L. Tang, D. Song, J.
Rhim, and Z. Chen, Fractal-like photonic lattices and localized states
arising from singular and nonsingular flatbands, APL Photon. \textbf{6},
116104 (2021).

\bibitem{PobleteAPX21} R. A. V. Poblete, Photonic flat band dynamics, Adv.
Phys. X \textbf{6}, 1878057 (2021).

\bibitem{FlachEL14} S. Flach, D. Leykam, J. D. Bodyfelt, P. Matthies, and A.
S. Desyatnikov, Detangling flat bands into Fano lattices, Europhys. Lett.
\textbf{105}, 30001 (2014).


\bibitem{GneitingPRB18} C. Gneiting, Z. Li, and F. Nori, Lifetime of
flatband states, Phys. Rev. B \textbf{98}, 134203 (2018).

\bibitem{ZSMPRB20} S. M. Zhang and L. Jin, Compact localized states and
localization dynamics in the dice lattice, Phys. Rev. B \textbf{102}, 054301
(2020).

\bibitem{ZSMPRA23} S. M. Zhang, H. S. Xu, and L. Jin, Tunable Aharonov-Bohm
cages through anti-$\mathcal{PT}$-symmetric imaginary couplings, Phys. Rev.
A \textbf{108}, 023518 (2023).

\bibitem{LeykamPRB17} D. Leykam, S. Flach, and Y. D. Chong, Flat bands in
lattices with non-Hermitian coupling, Phys. Rev. B \textbf{96}, 064305
(2017).

\bibitem{LiGePRA15} L. Ge, Parity-time symmetry in a flat-band system, Phys.
Rev. A \textbf{92}, 052103 (2015).

\bibitem{BlochPRL16} F. Baboux, L. Ge, T. Jacqmin, M. Biondi, E. Galopin, A.
Lema\^{\i}tre, L. L. Gratiet, I. Sagnes, S. Schmidt, H. E. T\"{u}reci, A.
Amo, and J. Bloch, Bosonic Condensation and Disorder-Induced Localization in
a Flat Band, Phys. Rev. Lett. \textbf{116}, 066402 (2016).

\bibitem{LiGePR18} L. Ge, Non-Hermitian lattices with a flat band and
polynomial power increase, Photon. Res. \textbf{6}, A10 (2018).

\bibitem{LQYuanAP22} G. Li, L. Wang, R. Ye, S. Liu, Y. Zheng, L. Yuan, X.
Chen, Observation of flat-band and band transition in the synthetic space,
Adv. Photon. \textbf{4}, 036002 (2022).

\bibitem{JoglekarPRA11} H. Vemuri, V. Vavilala, T. Bhamidipati, and Y. N.
Joglekar, Dynamics, disorder effects, and $\mathcal{PT}$-symmetry breaking
in waveguide lattices with localized eigenstates, Phys. Rev. A \textbf{84},
043826 (2011).

\bibitem{DavyNature22} M. Horodynski, M. K\"{u}hmayer, C. Ferise, S. Rotter,
and M. Davy, Anti-reflection structure for perfect transmission through
complex media, Nature \textbf{607}, 281 (2022).

\bibitem{LeykamPRB13} D. Leykam, S. Flach, O. Bahat-Treidel, A. S.
Desyatnikov, Flat band states: Disorder and nonlinearity, Phys. Rev. B
\textbf{88}, 224203 (2013).

\bibitem{HCWuPRL24} H. C. Wu, H. S. Xu, L. C. Xie, and L. Jin, Edge State,
Band Topology, and Time Boundary Effect in the Fine-Grained Categorization
of Chern Insulators, Phys. Rev. Lett. \textbf{132}, 083801 (2024).

\bibitem{JoglekarPRA10} Y. N. Joglekar, D. Scott, M. Babbey, and A. Saxena,
Robust and fragile $\mathcal{PT}$-symmetric phases in a tight-binding chain,
Phys. Rev. A \textbf{82}, 030103(R) (2010).

\bibitem{JoglekarPRAR11} Y. N. Joglekar and A. Saxena, Robust $\mathcal{PT}$%
-symmetric chain and properties of its Hermitian counterpart, Phys. Rev. A
\textbf{83}, 050101(R) (2011).

\bibitem{FleuryPRL19} F. Zangeneh-Nejad and R. Fleury, Topological Fano
Resonances, Phys. Rev. Lett. \textbf{122}, 014301 (2019).

\bibitem{KottosPRApp20} M. Reisner, D. H. Jeon, C. Schindler, H. Schomerus,
F. Mortessagne, U. Kuhl, and T. Kottos, Self-Shielded Topological Receiver
Protectors, Phys. Rev. Appl. \textbf{13}, 034067 (2020).

\bibitem{ClerkPRL23} M. Koppenh\"{o}fer, P. Groszkowski, and A.\thinspace A.
Clerk, Squeezed Superradiance Enables Robust Entanglement-Enhanced Metrology
Even with Highly Imperfect Readout, Phys. Rev. Lett. \textbf{131}, 060802
(2023).

\bibitem{SzameitNC20} M. Kremer, I. Petrides, E. Meyer, M. Heinrich, O.
Zilberberg, and A. Szameit, A square-root topological insulator with
non-quantized indices realized with photonic Aharonov-Bohm cages, Nat.
Commun. \textbf{11}, 907 (2020).

\bibitem{AhufingerLight20} C. J\"{o}rg, G. Queralt\'{o}, M. Kremer, G. Pelegr%
\'{\i}, J. Schulz, A. Szameit, G. V. Freymann, J. Mompart, and V. Ahufinger,
Artificial gauge field switching using orbital angular momentum modes in
optical waveguides, Light Sci. Appl. \textbf{9}, 150 (2020).

\bibitem{VicencioPRL22} G. C\'{a}ceres-Aravena, D. Guzm\'{a}n-Silva, I.
Salinas, and R. A. Vicencio, Controlled Transport Based on Multiorbital
Aharonov-Bohm Photonic Caging, Phys. Rev. Lett. \textbf{128}, 256602 (2022).

\bibitem{LeykamPRA12} D. Leykam, O. Bahat-Treidel, and A. S. Desyatnikov,
Pseudospin and nonlinear conical diffraction in Lieb lattices, Phys. Rev. A
\textbf{86}, 031805(R) (2012).

\bibitem{ThomsonPRL15} S. Mukherjee, A. Spracklen, D. Choudhury, N. Goldman,
P. \"{O}hberg, E. Andersson, and R. R. Thomson, Observation of a Localized
Flat-Band State in a Photonic Lieb Lattice, Phys. Rev. Lett. \textbf{114},
245504 (2015).

\bibitem{VicencioPRL15} R. A. Vicencio, C. Cantillano, L. Morales-Inostroza,
B. Real, C. Mej\'{\i}a-Cort\'{e}s, S. Weimann, A. Szameit, and M. I. Molina,
Observation of Localized States in Lieb Photonic Lattices, Phys. Rev. Lett.
\textbf{114}, 245503 (2015).

\bibitem{MolinaPRA15} M. I. Molina, Flat bands and $\mathcal{PT}$ symmetry
in quasi-one-dimensional lattices, Phys. Rev. A \textbf{92}, 063813 (2015).

\bibitem{RamezaniPRA17} H. Ramezani, Non-Hermiticity-induced flat band,
Phys. Rev. A \textbf{96}, 011802(R) (2017).

\bibitem{LGePRL18} B. Qi, L. Zhang, and L. Ge, Defect States Emerging from a
Non-Hermitian Flatband of Photonic Zero Modes, Phys. Rev. Lett. \textbf{120}%
, 093901 (2018).

\bibitem{ASzameitPRL19} T. Biesenthal, M. Kremer, M. Heinrich, and A.
Szameit, Experimental Realization of $\mathcal{PT}$-Symmetric Flat Bands,
Phys. Rev. Lett. \textbf{123}, 183601 (2019).

\bibitem{LJinPRA19} L. Jin, Flat band induced by the interplay of synthetic
magnetic flux and non-Hermiticity, Phys. Rev. A \textbf{99}, 033810 (2019).

\bibitem{SMZhangPRA19} S. M. Zhang and L. Jin, Flat band in two-dimensional
non-Hermitian optical lattices, Phys. Rev. A \textbf{100}, 043808 (2019).

\bibitem{SMZhangPRR20} S. M. Zhang and L. Jin, Localization in non-Hermitian
asymmetric rhombic lattice, Phys. Rev. Res. \textbf{2}, 033127 (2020).

\bibitem{AAndreanovPRB21} W. Maimaiti and A. Andreanov, Non-Hermitian
flat-band generator in one dimension, Phys. Rev. B \textbf{104}, 035115
(2021).

\bibitem{LChenAP23} B.-C. Xu, B.-Y. Xie, L.-H. Xu, M. Deng, W. Chen, H. Wei,
F. Dong, J. Wang, C.-W. Qiu, S. Zhang, and L. Chen, Topological Landau-Zener
nanophotonic circuits, Adv. Photonics \textbf{5}, 036005 (2023).
\end{thebibliography}
\end{document}